# Computing spatial information from Fourier coefficient distributions

William F. Heinz[*,1], Jeffrey L. Werbin[*,1], Eaton Lattman[2,3] and Jan H. Hoh[1,#]

[1]Department of Physiology, Johns Hopkins School of Medicine, Baltimore, Maryland

[2]Department of Biophysics, Johns Hopkins University, Baltimore, Maryland

[3]Present Address: Hauptman-Woodward Medical Research Institute, Buffalo, New York.

**Abstract**

We present an approach to computing spatial information based on Fourier coefficient distributions. The Fourier transform (FT) of an image contains a complete description of the image, and the values of the FT coefficients are uniquely associated with that image. For an image where the distribution of pixels is uncorrelated, the FT coefficients are normally distributed and uncorrelated. Further, the probability distribution for the FT coefficients of such an image can readily be obtained by Parseval's theorem. We take advantage of these properties to compute the spatial information in an image by determining the probability of each coefficient (both real and imaginary parts) in the FT, then using the Shannon formalism to calculate information. By using the probability distribution obtained from Parseval's theorem, an effective distance from the completely uncorrelated or most uncertain case is obtained. The resulting quantity is an information computed in k-space (kSI). This approach provides a robust, facile and highly flexible framework for quantifying spatial information in images and other types of data (of arbitrary dimensions). The kSI metric is tested on a 2D Ising ferromagnet, and the temperature-dependent phase transition is accurately determined from the spatial information in configurations of the system.

[*]These authors contributed equally.

[#]To whom correspondence should be addressed:

    Department of Physiology
    Johns Hopkins School of Medicine
    725 N. Wolfe Streets
    Baltimore, MD 21205

**Introduction**

The amount of information in a system that varies in time or space, for example, is an interesting and useful quantity. Information theoretical approaches are employed to study questions in fields ranging from linguistics [1] to chemistry [2] and cosmology [3]. In biology and related sciences information theory has been of long-standing interest [4]. Indeed, information would appear to be fundamental to widely studied problems such as cell-cell communication and signal transduction, as well as the genetic encoding and expression of proteins. However, while information theory has been applied to these [5,6,7,8] and other problems including directional cell movement [9], protein structure [10] and neuronal signal processing [11], to date it has had a modest impact in biology and biochemistry.

Our interest in this issue originates in a desire to understand spatial information in cellular microenvironments. For example, the extracellular matrix is a complex fibrous material composed of many different proteins that provides important signals that activate specific cellular functions [12]. We are attempting to understand how cells process spatial information in natural microenvironments, in part by using *in vitro* cell culture studies in which information is experimentally controlled [13]. However, the lack of a suitable information metric for those studies prompted us to develop an information measure based on correlations between structures or events in a system. This new metric is general and useful for any type of data, and we present the underlying theory as well several tests of the metric. Applications to biological problems will appear at a later time. Images are used as a vehicle for presentation, but the metric can be applied to any type of data for which a Fourier Transform can be obtained.

There is no single formal definition of "spatial information", but we use that expression in the conventional Shannon sense of observer-independent information in which information is related to the likelihood of an event [14, 15]. Information in this context is closely related to complexity and randomness [16]. A number of approaches to computing image information have been developed, many based on the well-known formalism established by Shannon in which

$$I_e = -k \log P_e \qquad (1)$$

and

$$H = -k \sum_e P_e \log P_e, \qquad (2)$$

where $P_e$ is the probability of the event e. Here $I_e$ is the information associated with the event *e*, and $H$ is the informational entropy associated with all possible events. When k = 1 and the log is base 2, the units are bits. Extending this to a collection of *i* elements, such as an image, *I* and *H* are both summed for all elements so

$$I_{total} = -\sum_i \log_2 P_{i,e} \qquad (3)$$

and

$$H_{total} = -\sum_i \sum_e P_{i,e} \log_2 P_{i,e}. \qquad (4)$$

In the most direct implementations of information theory, $P_e$ is computed from the statistical properties of the elements or events in the system. For example, in image analysis a common approach is to compute an "image entropy" ($H_{RS}$, subscript RS is for real space) by taking the value of each pixel as an event and using the normalized image histogram to determine $P_e$ [17]. However, such an approach does not capture the spatial organization -- that is, the spatial correlations between pixels -- within the image. The histogram is essentially a measure of information based on composition, where all images with the same histogram and the same number of pixels will have the same image entropy, as well as the same information ($I_{RS}$), no matter the arrangement of the pixels.

A number of solutions to capture spatial components of the information in data have been developed. Shannon's original solution to dealing with contributions from correlations between elements in the data was to group elements into n-grams [14]. The probabilities for the n-grams were then used to compute the information. While this can work well for small group sizes, it becomes increasingly difficult to employ as the n-gram size grows. A related approach was used for images to improve the histogram-based entropy measure of information; pair-wise pixels or voxels were used to compute a quantity that incorporates a spatial component [18]. This approach was subsequently extended to additional dimensions [19]. However, while the approach was shown to improve robustness of image registration, it is computationally intensive because the number of dimensions in the problem grows linearly with the radius of the neighborhood considered and the computational requirements scale with the number of pixels in the image raised to the number of dimensions. Thus only neighborhoods of a few pixels in diameter were used. A different approach based on the distribution of curvatures in 2- or 3-dimensional objects to reflect shape information has also been proposed [20]. With this metric, simple objects with constant curvatures, such as planes and spheres, have a shape information of 0. While objects with changes in curvature have more shape information.

A measure of spatial information also emerges from considering the algorithmic or Kolmogorov complexity, where the information is directly related to the size of smallest set of instructions that completely describe that object [21]. While it is not possible to compute the smallest set of instructions for an arbitrary image, the use of compression algorithms has become a practical implementation based on the same underlying idea [22]. The more compressible an image is – the less information it has.

Here we propose a new approach to quantifying spatial information that is based on an analysis of the Fourier coefficients computed from an image. By using Parseval's theorem to determine the expected variances for the distributions of coefficients, a Shannon-like quantity can be readily computed from the Fourier transform (FT) of any image. Because this information metric is computed in k-space, we refer to it as k-Space Information (kSI). The term k-space information has also been used in the analysis entropy associate with a particular pixel across a set of related images, although there a different approach is used to addresses a different problem than here [23]. We validate kSI in part by comparing it to the Kolmogorov complexity as approximated by JPEG2000 (JP2K) compression (lossless implementation).

**Theory and Computation**

We begin with an image $f(x,y) = \{f_{xy}\}$ where x and y are integers that index a pixel value, $f_{xy}$. The discrete Fourier transform of $f(x,y)$ is $F(m,n)$, where $F(m,n) = \{F_{mn}\}$, and $F_{mn}$ are complex numbers $F_{mn} = a_{mn} + ib_{mn}$. We note that the FT is an equivalent representation of the information in an image, and the inverse transform returns the original image. Further, any set of Fourier coefficients is uniquely associated with an image. In our approach, the probability distribution $P(e) = \{p_e\}$, where $\Sigma p_e = 1$, is the distribution of values for the Fourier coefficients, and the event e is a coefficient, F, having a particular value, a+ib. For a given coefficient $F_{mn}$, $P_{mn}(a,b) = \{p_{ab}\}_{mn}$ is the probability distribution of the real and imaginary parts of the coefficients $F_{mn}$, and $\Sigma\{p_{ab}\}_{mn} = 1$. For the case where the pixels in the original image are uncorrelated, or completely randomized, we assume all coefficients are independent. That allows us, by the application of the central limit theorem, to determine $P_{mn}(a,b)$, and establish that $P_{mn}(a,b) = P(a,b)$, $\{p_{ab}\}_{mn} = \{p_{ab}\}$ and that $\{p_{ab}\}$ obeys a normal distribution. Thus the entropy of a coefficient can be written as $H_{mn} = H = -\Sigma p_{ab} \log p_{ab}$, and therefore $H_{total} = N \cdot H$ (where N=the number of pixels in the image). Note that we employ the information theory convention that $p_{ab}\log p_{ab} = 0$ if $p_{ab} = 0$. The real and imaginary parts of the FT are also independent, and consequently $P(a,b) = P(a)P(b)$ and $H = -\Sigma p_a p_b \log p_a p_b$ where $\Sigma p_a = 1$ and $\Sigma p_b = 1$. The distribution of Fourier coefficients of independent random variables obeys a normal distribution with a mean of 0, so the probability density function (pdf) is

$$p_a p_b = \frac{1}{2\pi\sigma_a\sigma_b} e^{-\left(\frac{a^2}{2\sigma_a^2} + \frac{b^2}{2\sigma_b^2}\right)} \qquad (5)$$

$$= \frac{1}{2\pi\sigma_a^2} e^{-\left(\frac{a^2+b^2}{2\sigma_a^2}\right)} \qquad (6)$$

and

$$\sigma_a^2 = \langle a_{mn}^2 \rangle = \frac{1}{2}\langle F_{mn}^2 \rangle, \qquad (7)$$

where $\sigma_a^2$ and $\sigma_b^2$ are the variances for the distributions of the real and imaginary parts of the Fourier coefficients [24]. These variances can be calculated readily from the original real space image using Parseval's theorem, which relates the image to its Fourier transform by

$$\sum_{x,y} f_{xy}^2 = C\sum_{m,n} F_{mn}^2 = CN\langle F^2 \rangle = 2CN\sigma_a^2 \qquad (8)$$

where C is a normalization factor that is dependent on the specific implementation of the FFT, mostly commonly either 1 or N. In our programs C is equal to N. The underlying assumptions here were confirmed numerically [25].

To determine the Fourier coefficient-based information, $I_{kS}$, or Fourier coefficient-based entropy ($H_{kS}$) of a particular image, we begin by computing the Fourier transform of the image and recording the real and imaginary parts of the coefficients (FIG. 1). We then use Parseval's theorem to compute the variances for the distributions of the real and imaginary parts of the coefficients that would be expected from completely randomized images with the same histogram as the image of interest. These variances are then used to compute the probabilities, $P_a$ and $P_b$, for each of the coefficients in the FT of the image by integrating the pdf over a window from $c-0.005\sigma$ to $c+0.005\sigma$ (where c is either a or b) [25]. The probabilities are then used to compute $I_{kS}$ by summing $-\log P$ over all coefficients. $H_{kS}$ is computed by integrating the pdf in steps of $0.01\sigma$.

Here it is important to appreciate that the Fourier coefficients for many or even most images of interest are not independent and not normally distributed. Thus, what is being computed is the probability of a particular set of coefficients associated with the image of interest occurring in the FT of a random image. Note that the $H_{kS}$ can be computed from the image histogram alone, the computation does not require the coefficients from the Fourier transform, and the $H_{kS}$ of every coefficent is the same. Thus any two images that share the same image histogram have the same $H_{kS}$. But $I_{kS}$ depends on the specific set of coefficients in an image, making it suitable for quantifying the spatial information in the image. In addition to the $I_{kS}$, it is useful to think about the spatial information in a particular image relative to the greatest possible spatial information for an image with the same histogram. The $H_{kS}$ sets an upper bound for the $I_{kS}$, and we define the k-Space Information as the information in an image relative to the entropy,

$$kSI \equiv H_{kS} - I_{kS} = N_{pixels}\left(\sum_{\forall c} P_c \log_2 P_c\right) + \sum_m \sum_n \log_2 P_{c_{mn}}. \qquad (9)$$

Information is defined as a decrease or reduction of uncertainty [26]. We take the $H_{ks}$ to define the most random or uncertain distribution of Fourier coefficients. Thus the kSI provides a measure of information by virtue of being related to how far from random a particular collection of Fourier coefficients is.

**Results**
Fourier based entropy and information in 2D grayscale images
   To characterize the kSI approach we first examine 2D grayscale images where each pixel has an 8-bit intensity value. We begin with three test images that share the same histogram, but in which the pixels are organized differently: an image of a flower, an image where the pixels are organized from the lowest value to the highest value, and an image where the pixels are shuffled randomly in the xy plane (FIG. 2). Because these all have the same histogram, the $H_{RS}$ and $I_{RS}$ are identical. Similarly, $H_{kS}$ provides an upper bound for $I_{kS,}$ and it is identical for all three images. However, the $I_{kS}$ depends on the specific set of Fourier coefficients in each image and thus varies significantly. The ordered image has the lowest $I_{kS}$, the randomly shuffled image has the highest $I_{kS}$ and the original image has an intermediate $I_{kS}$. Subtracting the image $I_{kS}$ from the $H_{kS}$ gives the kSI for each image, showing that the ordered image is furthest away from random, and the flower image is between random and perfectly ordered. A qualitatively similar result is obtained using JP2K compression although the scale is inverted such that high complexity has large values and low complexity has low values.

Information varies with "disorder"
   One of the most basic qualifications for a measure of spatial information is that it will vary in a predictable way with the amount of order in an image. To test the behavior of the information metric developed here on disorder/order, we use a series of images where the amount of spatial information is changed in a well-defined fashion. Starting with an image in which the pixels are perfectly ordered (as in FIG. 2B, except from a uniform histogram), a varying fraction (from 0-1.0) of the pixels are shuffled to random positions (FIG. 3A). For this data set, the $I_{kS}$ increases monotonically with increasing amount of shuffle or disorder (FIG. 3B). Likewise the JP2K analysis shows a monotonic increase with increasing shuffle, although the rise is steeper at small shuffles and becomes less sensitive at higher shuffles. We have chosen to define kSI as the distance from a completely disordered image as defined by the $H_{kS}$. Thus the more order that is introduced into the image, the less information and the higher the kSI value (FIG. 3C). Note that the shuffled pixels are simply moved from one place to another, and thus the histogram is maintained. Interestingly, both the $I_{kS}$ and the kSI have a simple quadratic dependence on the shuffle fraction.

Precision estimate
   It is not possible to produce images with a known spatial information, and thus not possible to independently establish the accuracy of the kSI metric. However, it is possible to estimate the precision of the metric. Because the kSI introduces an arbitrary offset to the spatial information, we perform this analysis on the $I_{kS}$. We start with a grayscale image and make a large number of images that are shuffled by a specific fraction. We then examine the distribution of $I_{kS}$ values for these shuffled images (FIG. 4). These values are approximately normally distributed. The variance of the distribution also depends on the amount of order in the image, and it is convenient to represent it as a percentage of the information. Thus for a 0.5 shuffle fraction the standard deviation is 0.045% while for an 0.85 shuffle fraction it is 0.014%. The variance of the coefficient distribution provides an upper bounds to the precision of the measurement; the precision cannot be any worse than that reflected by the variation that arises from repeated shuffles. Instead, some of the variance from the repeated shuffling will likely produce images that vary significantly in kSI – thus broadening the distribution beyond that defined by the precision. The JP2K metric is qualitatively similar to the kSI with a similar precision (0.054% standard deviation for an 0.5 shuffle, and 0.040% for an 0.85 shuffle).

One other effect on the precision arises from the fact that there is a small difference in the kSI that is computed for images with different rotations. When a rectangular image is rotated an integer multiple of 90 degrees the spatial information is clearly conserved, and one would expect the kSI for all such rotations to be identical. However, the kSI for rotations vary slightly [25]. This difference arises because of a shift of indices upon rotation of the image. The rotation-related error is small and we typically ignore it. If needed, the index shift can be eliminated by padding the first row and the first column of an image with zeros [25] although the padding itself introduces information into the image.

Composition dependence of the information

Two-dimensional images are typically intensity-based representations with variable histograms. For these types of images the $I_{kS}$ and kSI are sensitive to the composition (*i.e.* histogram) of the image. This is an expected result from Parseval's theorem, since the variance depends explicitly on the composition. The $H_{kS}$ on the other hand is composition independent. This somewhat counter intuitive result is related to the discrete computation of $H_{kS}$. When the integration window for the analysis is set to a fraction of the variance, the $H_{kS}$ becomes identical for all images of the same size and the kSI for the 1.0 shuffled images approaches zero (irrespective of the histogram).

A histogram independent surface-based approach

One of our main interests here is to develop a metric that can be applied to a broad range of systems and that readily allows for comparison of spatial information between systems. The histogram dependence described for the 2D grayscale images above complicates comparisons. To solve this problem we recast a 2D grayscale image into a surface in a binary n+1 dimensional structure, where the extra dimension is a binary representation of the intensity value at each position in the original structure (FIG. 5). For an 8-bit grayscale image with N pixels the additional dimension has 256 elements, such that a grayscale image is effectively recoded into a 3D surface. We adopt a nomenclature for the spatial information, (u:w)D, where u is the number of spatial dimensions and w is the number of additional dimensions that are used to encode non-spatial properties. Thus the information and entropy computed for a 2D grayscale image is (2:0)D, for the 3D representation these values are (2:1)D. Now, all images of the same size will have identical histograms: N values of 1, and 255*N values of 0. Thus all images of the same size have the same variance by Parseval's theorem, eliminating the histogram effect when comparing images. The approach is in effect similar to the symbolic vectorization of sequence data that has been used for DNA analysis [27, 28].

Aside from the histogram effects, the binary surface approach for the kSI metric qualitatively recapitulates all the results from the (2:0)D approach described above (FIG. 6). As with the (2:0)D case the $H_{kS}$ gives the largest possible value for $I_{kS}$, and the kSI decreases monotonically with increasing shuffle fraction. It should be noted that for the (2:1)D analysis we use an original image that is a flat plane where all the z values are the same, and that the shuffling occurs in three dimensions. The shuffle series again shows a quadratic relationship between kSI and shuffle fraction. Interestingly, in this case the JP2K metric increases from the most ordered image to an 0.7 shuffle, but then decreases slightly up to 1.0 shuffle. One notable difference between the (2:0)D and (2:1)D approach is that the precision for the (2:1)D approach is significantly better. For example, for the 0.5 and 0.85 shuffled images, the (2:1)D standard deviation is $4 \times 10^{-9}$% and $7 \times 10^{-9}$% respectively. While these computations are more memory intensive and require greater computational effort, we have found the (2:1)D approach for grayscale images significantly more useful than the (2:0)D.

Test of kSI metic on an Ising model

To provide a physical test of the kSI metric we examined the phase transition in a 2D Ising ferromagnet, for which there is an exact solution [29]. For this analysis energies and kSI's were computed for configurations at a series of temperatures (FIG. 7A and B). The energies for the system show the phase transition at a normalized temperature of 2.26+/-0.01. From the kSI based analysis the transition temperature was also 2.26+/-0.01 [25]. These values agree to within error with the known transition temperature of 2.27 [29]. Thus the phase transition of the Ising ferromagnet can be accurately determined from the spatial distribution of spins alone. We note that the energy and information in this analysis are close to linearly related over a large fraction of the temperature range, including the temperatures around the phase transition, but deviate from each other substantially at the highest temperatures.

**Discussion**

The main result presented here is an approach based on the Fourier transform to represent the spatial information content of a 2D image. Unlike approaches based on the Shannon formalism that use the distribution of pixel values as the probability distribution, this calculation takes into account the spatial arrangements in an image. The kSI metric provides a measure of how correlated the positions of pixels are relative to the limit where all the pixel positions are uncorrelated (as defined by the $H_{kS}$). It in effect describes an image by how different it is from a randomized image with the same histogram. The utility of this metric is illustrated by showing that it is sensitive to the phase transition in a 2D Ising model based only on the spatial distribution of spins.

Extending the kSI analysis of a 2D intensity based data set to a 3D binary representation makes the metric histogram independent and dramatically improves the precision the kSI measurements thus increasing the sensitivity significantly. Further it enables the direct comparison of kSI numbers for any images that are equal in size. Additional dimensions can also be used to encode other non-spatial information such as composition. Thus our ongoing work with the kSI metric has been almost exclusively with the (2:1)D approach.

Contributions from different spatial frequencies

We assume equal contributions to the kSI from the different spatial terms in the FT. However, it is likely that there will be reasons to use different weights, and specific weights might be employed to capture information at specific frequencies or combinations of frequencies. Such filtering of the contributions to the final sum, including setting some values to zero, is technically straightforward.

The basis set issue

A limitation of the FT approach to computing information is that the FT does not provide an optimal basis set for all images. This is effectively the same problem that plagues the Kolmogorov complexity. Thus, the FT approach provides an upper bound to the spatial information in an image. The extent to which that limits the utility of the FT approach is not easily determined, but it is a subject for further study.

Higher dimensions

In this paper we have described the treatment of a one-component image – that is an image with one type of element (a pixel) that varies only in intensity (the value of the pixel), as well as the projection of that image into three dimensions. However, the approach is readily extended to deal with higher (or lower) dimensional data as well as multiple components. In the case of multiple components, additional dimensions can be used to encode additional qualities of a pixel. The higher dimensions can represent any variable, atom type, time, temperature, connectivity *etc*. Such vectorization is as earlier noted commonly seen in bioinformatic analysis of DNA sequences [27, 28]. For example sequences of DNA can be represented as a string of numbers with an arbitrary scale such as, A =1, T=2, G=3 and C=4. However, a number assignment is typically arbitrary, and the results of an analysis depends on the particular assignment used in a meaningless way.  On the other hand, the base at each position in the sequence by a vector that has 4 elements that can have values of either 1 or 0 to represent the presence or absence of a base respectively at that position.  In this case the representation is independent of the assignment, and the entire sequence is described in a histogram-independent manner.  Extending that logic we can examine complex data structures with input from multiple sources by representing the different types of data in separate dimensions.  The Fourier transform can be readily applied to structures of many dimensions, permitting the kSI for almost any data set to be calculated.

**Conclusion**

The k-space-based approach to computing information presented here is a facile and general method that yields a metric for spatial information that provides the basis for further developments in many fields of study.  We also note that the k-space framework can be readily extended to almost any application in which information theory is currently used, such as computing joint or mutual information between images.  Information is used to characterize the complexity of physical systems [30, 31], including the biochemical and biological systems of interest to us, and the kSI approach lends itself to such applications.

**Acknowledgements**

The authors thank Dr. Thomas Woolf and Dr. David Haviland for interesting and helpful discussions, as well as Mr. Gardner Swan for programming assistance.

**Figure Legends**

FIG. 1. Schematic representation of the approach for computing entropy and information in k-space based on Fourier coefficient probabilities. To compute the entropy, Parseval's theorem is first used to obtain the variance for the Fourier coefficients based on the values of the pixels in the image. The Fourier coefficients are normally distributed, and thus the variance provides a probability density function. This probability density function is then divided into bins of $\sigma/100$, and the entropy is computed by summing $P_c Log_2 P_c$ from $-10\sigma$ to $+10\sigma$ and multiplying by twice the number of coefficients. Note that the $\forall c$ limit is used to indicate that the summation is being taken over all the bin values of a or b. To compute the spatial information, the real and imaginary parts of the Fourier coefficients for the image are determined. A probability for each coefficient is obtained from the probability density function above, and the information is computed by summing $Log_2 P$ for both parts.

FIG. 2. Information metrics for three test images with identical histograms but varying spatial organization. A. Image of a flower. B. Image with the same histogram as the flower, with the pixels arranged from highest to lowest values. C. Image with the same histogram as the flower, with the pixels arranged randomly. The units are bits for all quantities.

FIG. 3. Dependence of spatial information metrics on image order/disorder. A. Starting with a 256x256 pixel 8-bit image composed of pixels arranged from highest to lowest (with a uniform histogram), the amount of order was varied by shuffling increasing fractions of pixels. The ordered image is selected because it has one of the most ordered arrangements possible. Although it may not have the absolute lowest $I_{kS}$, it represents a low-information bound. B. For this shuffle series the $I_{kS}$ increases monotonically from the most ordered to the most disordered image ($I_{kS}$ = -942.1318254$x^2$ + 20550.8962979x + 1027065.0735181 and $R^2$>0.9999999, where x is the shuffle fraction). The JP2K compression also increases monotonically with increasing disorder. C. The kSI decreases monotonically with increasing disorder in an image (kSI = 94213.1828343$x^2$ - 186666.3268458x + 92458.6869676 and $R^2$>0.9999999).

FIG. 4. Precision estimate for $I_{kS}$. A. Distribution of the $I_{kS}$ values for 50% shuffles plotted as a deviation from the mean (1,116,445 bits). 10,000 uniform images with a uniform histogram were each randomly shuffled to 50%, and the $I_{kS}$ for each was computed. B. Distribution of JP2K values for the same 10,000 images as in A (mean=4,446,947 bits). C. Precision of the $I_{kS}$ and JP2K, represented by the %error (standard deviation/mean), as a function of shuffle fraction.

FIG. 5. Illustration of the mapping of a 2D grayscale image, in which (2:0)D metrics are computed, to a 3D binary surface in which (2:1)D metrics are computed. The grayscale and shading in the 3D representation is for presentation purposes only.

FIG. 6. (2:1)D spatial information analysis of a 3D shuffle series. This shuffle series begins with a 256x256x256 voxel 1-bit image where the value for z=0 at all xy points is 1 and all other values are zero. The shuffling is then carried out by randomly selecting an xy point, and and setting a random z value (voxel) to 1. A. (2:1)D $I_{kS}$ computed from the 3D (1-bit) representation and JP2K computed from the 2D (8-bit) representation as a function of shuffle fraction ($I_{kS}$ = -242336.26457$x^2$ + 5329091.80210x + 262319053.10302 and R2>0.99999)**.** The JP2K value has minimum for the most ordered image, but reaches a maximum at a shuffle fraction of ~0.7. B. The kSI decreases monotonically with increasing disorder in an image. Also, the kSI fits the quadratic expression kSI = 242035.31665967$x^2$ - 5325011.46447971x + 29290264.54958770 and $R^2$>0.99999999). C. Precision of the (2:1)D kSI and JP2K, represented by the %error

(standard deviation/mean), as a function of shuffle fraction.

FIG 7. Information based analysis of the phase transition in a 2D Ising ferromagnet. A. Representative snapshots of system configuration at different temperatures. The simulations were performed using Ising 1.1 (written by Dr. D. Schroeder). Configurations were generated by first randomizing the spins, and then running the simulation for $5 \cdot 10^7$ steps for a 400x400 spin system (with periodic boundaries). The energy for each configuration was computed. The configurations were also converted to 8-bit gray scale TIFF files and the kSI was computed for each. B. The kSI shows a well defined phase transition at a temperature of 2.26+/-0.01 [25]. The well-known temperature dependent phase transition based the computed energies of the system is shown for comparison. C. Relationship between spatial information and energy in the 2D Ising ferromagnet.

Figure 1

$$H_{kS} = \sum_n \sum_m \left( \sum_{\forall a} P_a \log_2 P_a + \sum_{\forall b} P_b \log_2 P_b \right)$$

$$= N_{pixels} \left( 2 \sum_{\forall c} P_c \log_2 P_c \right)$$

$$I_{kS} = -\sum_m \sum_n \log_2 P_{a_{mn}} - \sum_m \sum_n \log_2 P_{b_{mn}}$$

Figure 2

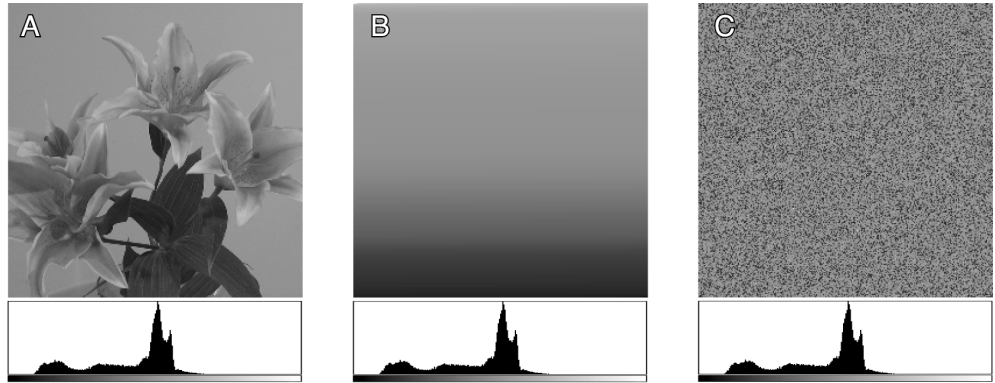

Figure 3

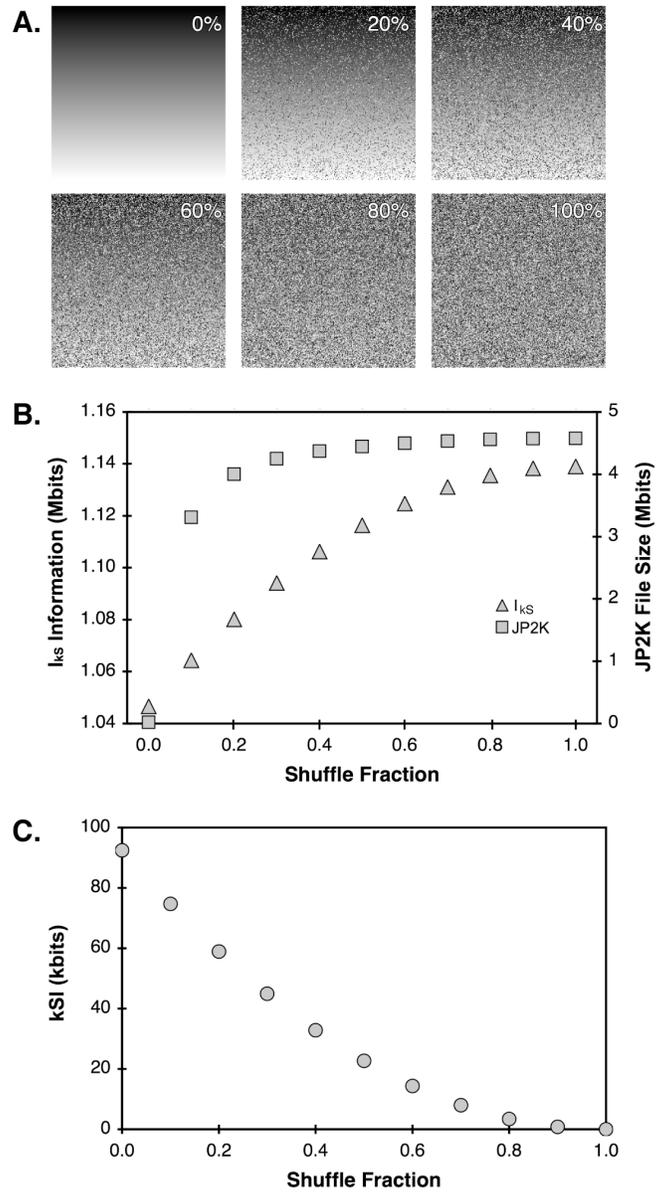

Figure 4

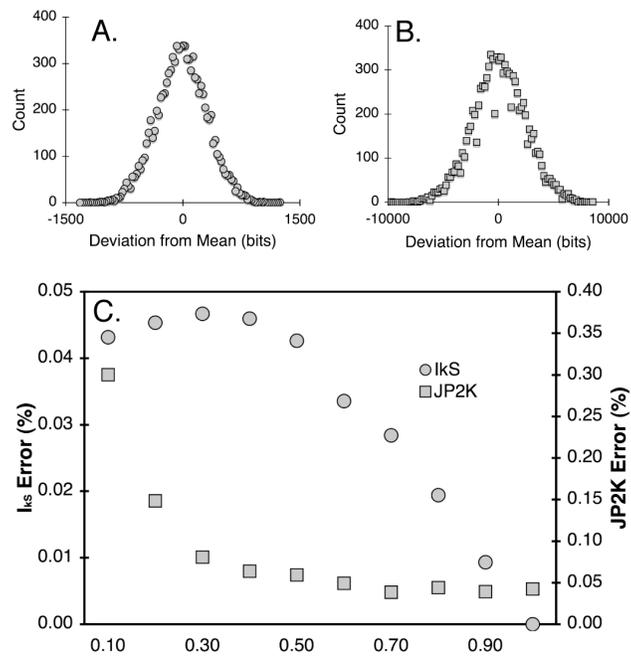

Figure 5

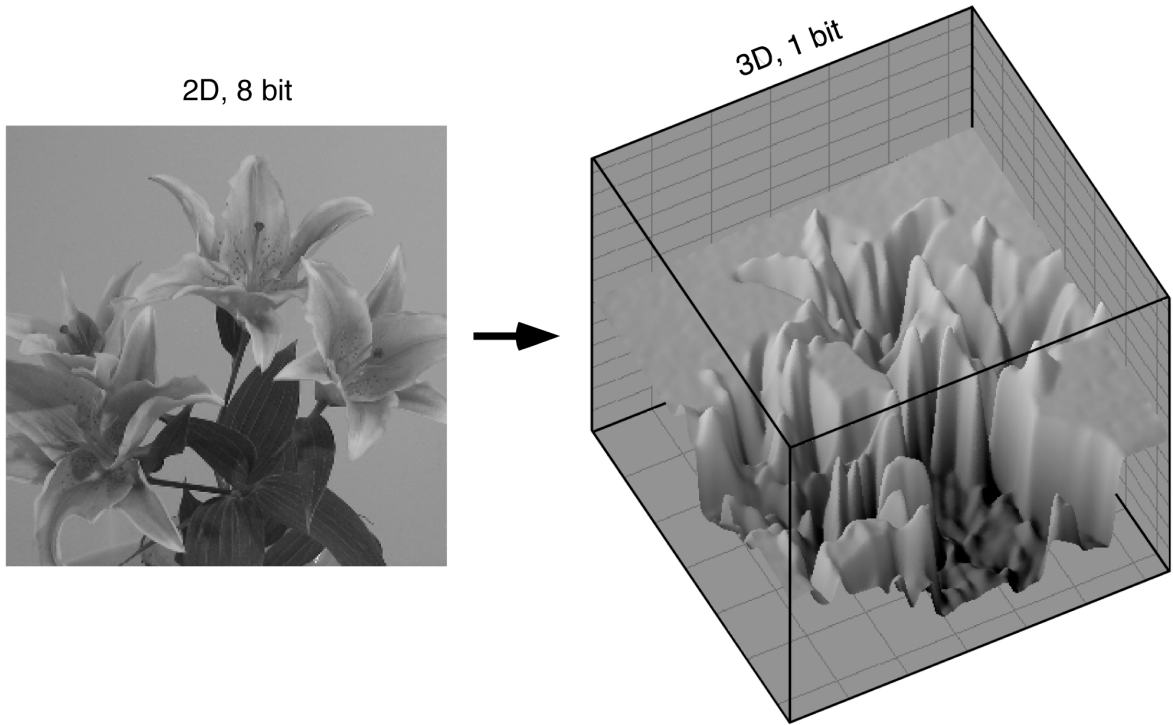

Figure 6

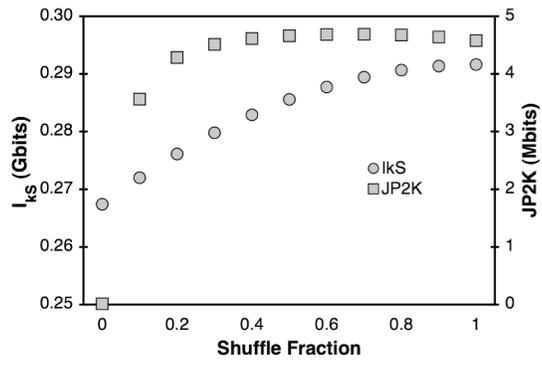

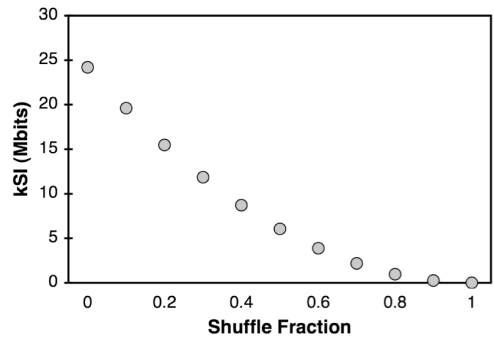

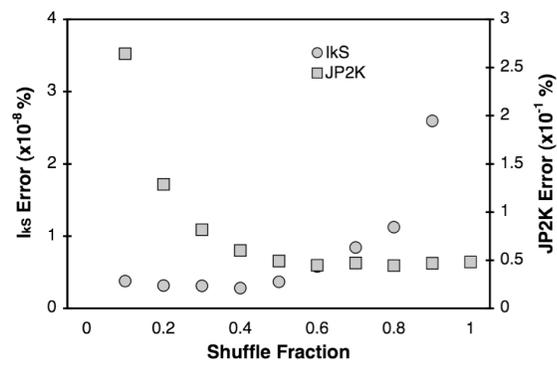

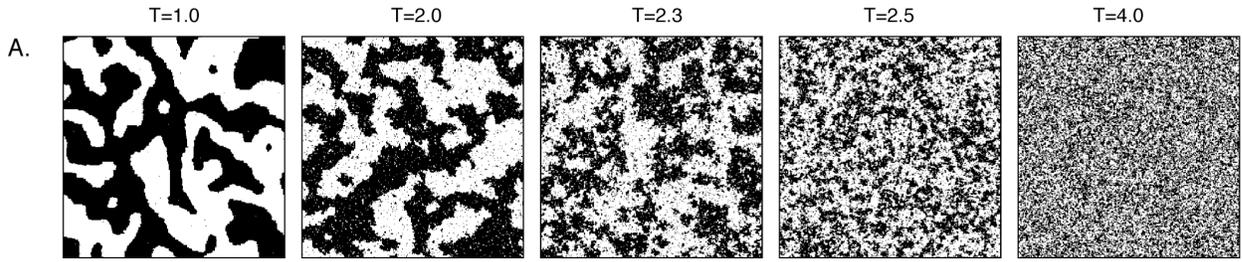

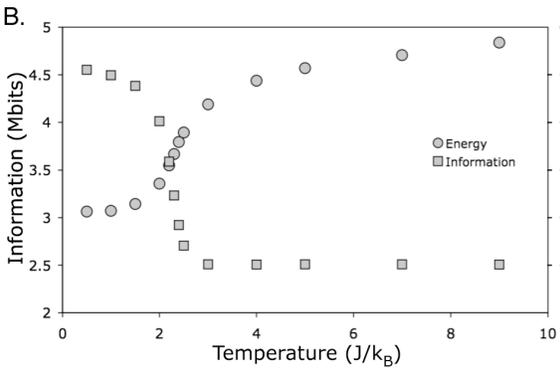
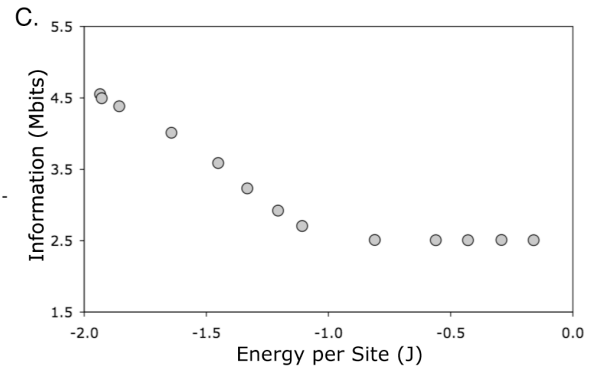

**Auxiliary Material**

Computational Verification of Fundamental Assumptions

Several assumptions are made in order to develop the information theory metric presented in this paper. In this section we computationally verify those assumptions.

*Assumption 1: The real and imaginary parts of the Fourier coefficients follow a normal distribution centered about zero for images where the pixel organization is not correlated.*

A 256x256 grayscale image with a uniform histogram was fully shuffled, so that all pixels where moved to a new position randomly, 10,000 separate times. Fourier transforms of each of the resulting images were computed, and the distributions of the real and imaginary parts of the coefficients were histogrammed separately. These distributions were normalized, and fit to a normal distribution (FIG. S1).

*Assumption 2: The distributions for the real and imaginary parts of the Fourier coefficients have the same variance ($\sigma^2$).*

From the data in assumption 1, we see that the variances for both the real and imaginary coefficients are identical to within the reported error of the fittings (FIG. S1 and TABLE S1).

*Assumption 3: The variances of the real and imaginary distributions are related to the histogram of the original image by Parseval's theorem.*

Given that the Fourier coefficients are normally distributed with a mean of 0 and a standard deviation of $\sigma$, Parseval's theorem shows that $\sigma$ can be calculated from the original image by the following equation.

$$f_i = \text{Image}_i - \overline{\text{Image}}$$

$$F = \mathcal{F}\{f\}$$

$$\sum_i |f_i|^2 = N_{pixels} \sum_i |F_i|^2 = N_{pixels} \langle |F|^2 \rangle = 2 N_{pixels} \langle |F_{real}|^2 \rangle = 2 N_{pixels} \langle |F_{imaginary}|^2 \rangle = 2 N_{pixels}^2 \sigma^2 \quad \text{(S1)}$$

$$\therefore \sigma = \sqrt{\frac{\sum_i |f_i|^2}{2 N_{pixels}^2}} = \sqrt{\frac{\sum_{\forall f} h(f)|f|^2}{2 N_{pixels}^2}}$$

where f is the normalized original image effectively removing the DC offset, $\mathcal{F}\{\}$ is the Fourier transform and $h(f)$ is the normalized histogram of the original image.

We calculated $\sigma$ by fitting the data from assumption 1 to a normal distribution and compared it to the $\sigma$ from Parseval's theorem. Both are identical within the error of the fit (Table S1).

One consequence of Parseval's is that $\sigma$ is dependent on the histogram of the real space image (eq. S1). Figure S1B shows data for the distribution of coefficients from an image with the same real space histogram as the flower image in Figure 2A. Both of these sets of images have normally distributed Fourier coefficients but different $\sigma$'s. For both sets of images the fitted $\sigma$'s and Parseval's $\sigma$'s are identical to within error (Table S1), establishing that Parseval's theorem can be used to accurately calculate the $\sigma$ of the Fourier coefficients from the real space image.

| Histogram Type | $\sigma_{real}$ from Fitting | $\sigma_{imaginary}$ from Fitting | $\sigma$ from Parseval's eq. |
|---|---|---|---|
| Uniform | 0.20415±2.51e-05 | 0.20410±2.55-05 | 0.20412259 |
| Flower | 0.10117±1.25e-05 | 0.10116±1.26e-05 | 0.10116563 |

Table S1. Comparison of standard deviations of the Fourier coefficients obtained computationally with those obtained analytically from Parseval's theorem. One image has a uniform histogram, and the other has a histogram that identical to that of the Flower image in Figure 2.

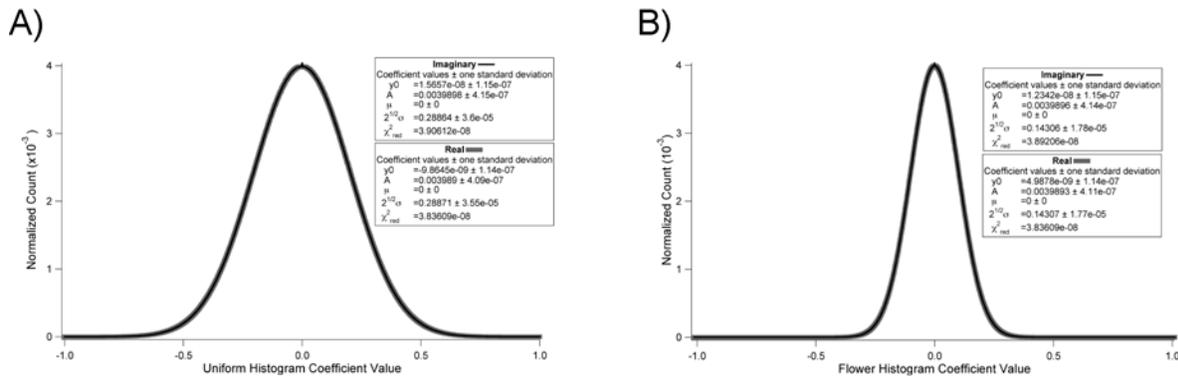

FIG. S1. Normalized Fourier coefficient histograms made from 10,000 shuffled images. The real parts are shown in thick grey and the imaginary parts are shown in thin black. All curves were fit to a normal distribution function centered at zero and the fitting parameters are reported in the graphs. The fits themselves are not shown but are indistinguishable from the data. A) From images with a uniform histogram. $\sigma_a$ = 0.20415±2.51•$10^{-5}$ and $\sigma_b$ = 0.20410±2.55•$10^{-5}$. B) From images with the same histogram as the flower image in Figure 2 $\sigma_a$ = 0.10117±1.25•$10^{-5}$ and $\sigma_b$ = 0.10116±1.26•$10^{-5}$.

Detailed example

To illustrate more explicitly how the $H_{kS}$ and the $I_{kS}$ are computed, we show the full calculation for a 3x3 pixel image (with three gray values). This is a (2:0)D calculation.

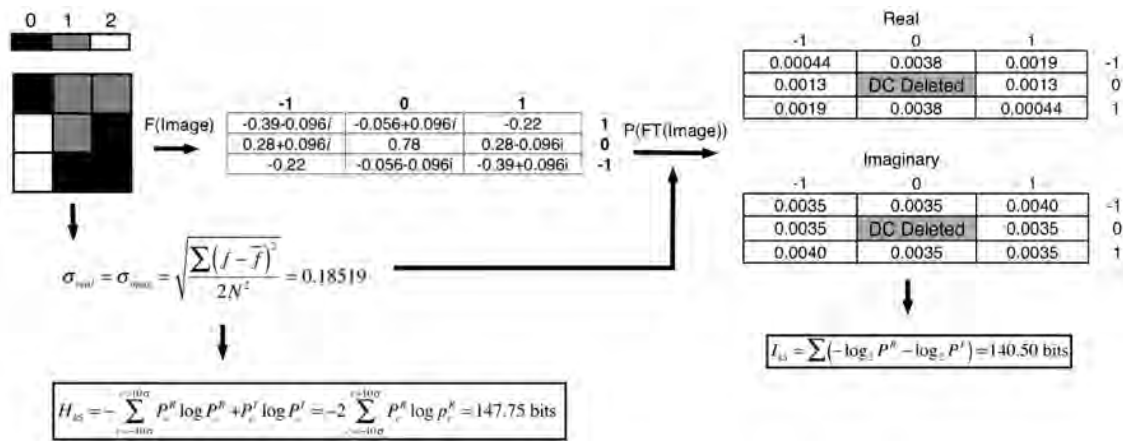

FIG. S2. (2:0)D $H_{kS}$ and $I_{kS}$ calculation for a 3x3 image. Note that the spatial information values computed for very small images such as this are not well behaved, and the example is for illustrative purposes only. We typically use a minimum image size of 8x8 pixels.

Effect of integration window on information computation

The normal distribution determined by Parseval's theorem is the probability density function (pdf), and to find $P^{image}(e)$ one must integrate the pdf.

$$P_e^{image} = \int_{e-\delta/2}^{e+\delta/2} \frac{1}{2\pi\sigma^2} e^{-\frac{x^2}{2\sigma^2}} dx \quad (S2)$$

The calculation of $P$ is sensitive to the size of the integration window, $\delta$. We have examined the effect of window size, and find that for computing kSI a window of $\delta = \sigma/100$ is suitable for most applications. At window sizes less than $\sigma/100$ there is a close to logarithmic dependence between kSI and the window size (FIG. S3). At window sizes of $\sigma/10$ and greater there is a deviation from that relationship.

For the calculation of $H_{kS}$ we also discretize the events with a window of $\delta$, from $-10\sigma$ to $10\sigma$. H of the normal distribution is known for continuous space, but this value can be negative. This is odd given that I is always greater than zero and H is the expectation value of I. To avoid confusion we discretize and take the Riemann sum of Plog P to determine H.

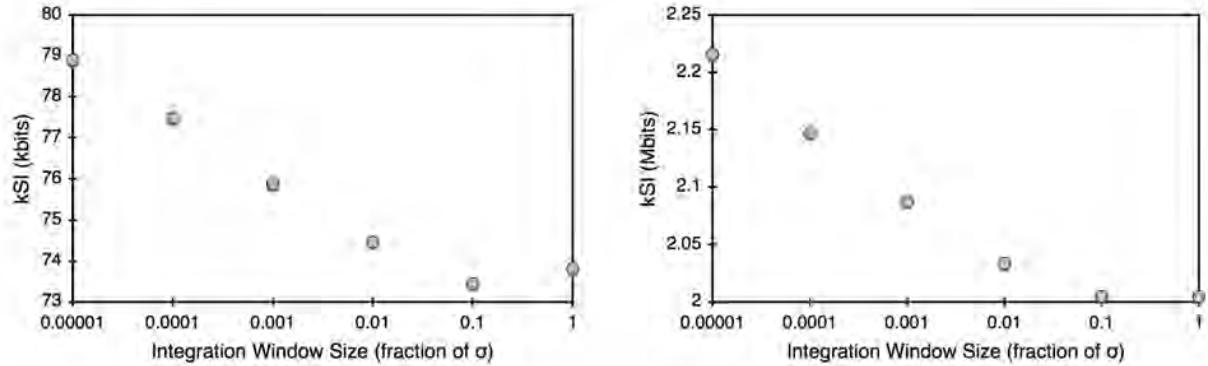

FIG. S3. Effect of integration window size on the computed (2:0)D and (2:1)D kSI values for a flower image. This relationship holds for the entire range over which it has been examined (square images from 8x8 to 512x512 pixels).

Image Rotation Error

To establish the effect of image rotation, we performed a shuffle series of an image with a uniform histogram of the type shown in Figure 4. Each of the images in the shuffle series was then rotated 90, 180 and 270 degree. The deviation of each of the rotations from the average (of the four rotations) was determined, and plotted as a function of shuffle fraction (FIG. S4). Because the effect of rotation is small we typically ignore it. Although, because the error arises from a shift in indices upon image rotation, the rotation dependence of the kSI can be eliminated by padding the first row and first column of the image with zeroes. (TABLE. S2).

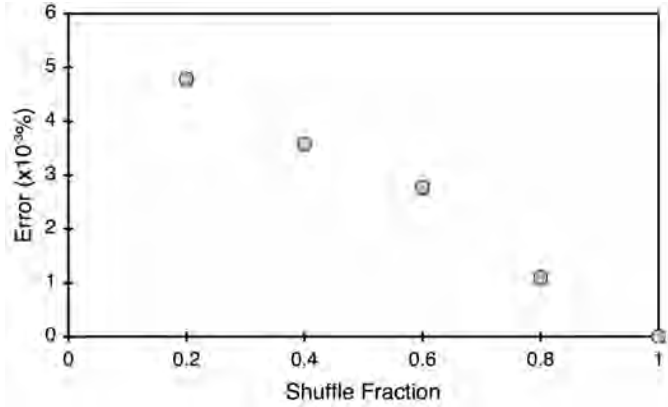

FIG. S4. Deviations in $I_{kS}$ that result from image rotation. Analysis of 100 images at each shuffle fraction (400 images after rotation).

|  | Padded | | Unpadded | |
|---|---|---|---|---|
| Rotation | (2:0)D | (2:1)D | (2:0)D | (2:1)D |
| 0 | 271127.046925841 | 72516111.5849259 | 264334.809013918 | 71380310.6512577 |
| 90 | 271127.046925841 | 72516111.5849259 | 264565.317342388 | 71379692.9973799 |
| 180 | 271127.046925841 | 72516111.5849260 | 264485.274565517 | 71383127.7719158 |
| 270 | 271127.046925841 | 72516111.5849261 | 264238.739585776 | 71382686.1803473 |

TABLE S2. Eliminating rotation error by image padding. $I_{kS}$ for a flower image that is 127x127 pixels, with and without padding the first row and column with zeroes. Both (2:0)D and (3:0)D analyses were performed for each. Padding eliminates differences that arise from rotation.

Determination of the phase transition temperature in the 2D Ising Ferromagnet

To determine the phase transition temperature for the 2D Ising ferromaget simulation, the region around the transition in Figure 7 was examined in detail. The Ising (version 1.1) program was run starting from a random configuration, with periodic boundary conditions, 400x400 pixels, $1 \cdot 10^9$ steps, and 8-bit gray scale TIFF images were output. 6 independent series of configurations were collected at 0.01 temperature increments around 2.27, and the energies for each of these configurations recorded. The kSI was then computed for each configuration. The temperatures for the maximum derivatives for both the energy and kSI as a function of temperature were then averaged.